# P ≠ NP Proof

André Luiz Barbosa
http://www.andrebarbosa.eti.br
Non-commercial projects: SimuPLC – PLC Simulator & LCE – Electric Commands Language

***Abstract***. *This paper demonstrates that P ≠ NP. The way was to generalize the traditional definitions of the classes P and NP, to construct an artificial problem (a generalization to SAT: The XG-SAT, much more difficult than the former) and then to demonstrate that it is in NP but not in P (where the classes P and NP are generalized and called too simply P and NP in this paper, and then it is explained why the traditional classes P and NP should be fixed and replaced by these generalized ones into Theory of Computer Science). The demonstration consists of:*

1. *Definition of Restricted Type X Program*
2. *Definition of the Extended General Problem of Satisfiability of a Boolean Formula – XG-SAT*
3. *Generalization to classes P and NP*
4. *Demonstration that the XG-SAT is in NP*
5. *Demonstration that the XG-SAT is not in P*
6. *Demonstration that the Baker-Gill-Solovay Theorem does not refute the proof*
7. *Demonstration that the Razborov-Rudich Theorem does not refute the proof*
8. *Demonstration that the Aaronson-Wigderson Theorem does not refute the proof*

*The paper demonstrates three new revolutionary ideas in the foundations of Computational Complexity Theory, against the established theory in the area:*

1. *Language Incompleteness: There are computational decision problems that are not languages (cannot be modeled as string acceptance testing to languages);*

2. *NP-Completeness Incompleteness: The Cook-Levin Theorem and the concept of NP-Completeness are false, whence it cannot in general be applied on an instance of the XG-SAT to reduce it to one of the SAT; and*

3. *SAT's weakness: The complexity class of the SAT maybe does not decide P versus NP question (if SAT is not in P then NP ≠ P; on the other hand, SAT in P doesn't imply NP = P).*

*The objective of the paper is just help the complexity theorists about their main question: If I am right, then the P versus NP Clay Mathematics Institute Prize must be canceled and replaced by the correct problem intended by them:* ***"Is SAT in P?"***

## Contents

## 1. Introduction

I am a very ambitious person: I have restated the P versus NP question on my own terms, and this fact is confused with a supposed lack of rigor in the paper. But this restating is not arbitrary or nonsense thing. I have not invented a new one or distorted the problem: I have just generalized it in order to hold all *computational decision problems* that can be constructed in the real world – the set of the $L_z$*-languages* –, not only the set of the *languages*, a small proper subset of the former, as stated in Section 3.3.1.

I have succeeded where better Math warriors in the area failed before me because I have seen that the Complexity Theory was incomplete, treating *languages* as if they would represent all the *computational decision problems*. Then, to solve the problem, I simply have generalized the theory in order to really include into it all the possible *decision problems* that can be decided by computers, not only the *languages*. (See Section 3.4.)

Hence, as Mathematics loves generalizations (since the concepts, notation, generality, power and applicability of the more general mathematical constructs are clearer and stronger than of the less general ones), I think I should not be considered so crank or crackpot by this "heresy". Here, in its generalized form, the profoundest question in Computational Complexity Theory is settled – and, in addition, new and marvelous paths for original and relevant researches are glanced; like the generalization in these terms of all Computation Theory (we can call it *The Barbosa's Program*). This *Program*, in fact, was already started in [13], with a $P \neq RP$ proof (leading to $P \neq BPP$, $P \neq ZPP$, and other great related results), in [19], with an $NP \not\subset P/poly$ *proof* (leading to $NEXP \not\subset P/poly$, $NEXP \neq MA$, $NEXP \neq BPP$, and other great related results), and in [15], with the wonder and astoundingly (wrong, unfortunately) conclusion that this Proof on my own terms leads to $P \neq NP$ on traditional ones (where the *uniformity polynomial time bounds blindness* in the technical report in [15] makes the errors in all its main conclusions, obliterating the fact that the *XG-SAT* is a single problem, even though its polynomial time bound is not *uniform* – see Def. 3.7, where this wide misconception is fixed). So, I so hope that the Computer Science community is ready to this stunning inflection point, originating deeper understanding on machine-based computation.

(**Note**: This paper of mine has a little non-technical comments, and its writing style is manifestly unusual, written in first person singular number at some places, which is not expected at all in technical works. But it is needed that I just do it in this way. I claim huge breakthroughs: If I am right, enormous stones will roll deeply into Science of Computation, whence this little paper can either be my eternal humiliation, or the glory of independent and fearless thinking, inspiring everlastingly the new generations of the Science thinkers. Thus, appreciating your precious time, I beg for your mercy and for your [unconventional] reading.)

Accordingly, in Sections 2 and 3 the *restricted type X programs* and the *XG-SAT* problem are formally defined, and some notes are included in order to avoid the traps in these definitions. In order to define the *XG-SAT*, computational decision problem and poly-time DTM are redefined in more general form, and then the Cook-Levin Theorem is disproved. So, it is proved that the *XG-SAT* is in NP, with concepts of poly-time *verifier* and *certificate of membership*. In Section 4, this demonstration is repeated with the old kind, using decider poly-time NTM. Then, in Section 5 it is proved that this problem is not in P (therefore, **P ≠ NP**, naturally), by demonstrating that it is impossible that any poly-time deterministic computation solves the *XG-SAT*.

In this proof, nothing is assumed about type, structure, form, code, nature, shape or kind of computation, neither structure (nor lack thereof) of data, eventually used into any DTM that tries to decide the problem in polynomial time. Otherwise, my proof exploits properties of computation that are specific to real world computers (without *oracles, infinite TMs* and other supernatural devices). In Sections 6, 7 and 8, it is demonstrated that the theoretical barriers against possible attempts to solve the P vs. NP question are not applicable to refute my proof. Finally, in Sections 8 and 9 there are some comments about related work (or lack thereof) to really solve this question, and references, respectively.

Shortly, in order for this P ≠ NP proof of mine be accepted, it is sufficient that the fact if there is an $L_z$-language (promise problem) separating complexity classes, then they are truly distinct, and the Def. 3.7 are both accepted. On scientific revolution/paradigm shifts, see [17].

Please: If possible, read the paper without preconceptions. Remember that, compared to proper Mathematics, immortality, fame, money and so on are vain things, silly concepts and bullshits that blow in the wind. The true happiness to an actual mathematician is to find the light into the darkest night. (By the way, see [25, 26, 27] for another amazing insights.)

## 2. Definition of Restricted Type X Program

**Definition 2.1.** Let **S** be a deterministic computer program, let **n** be a finite positive integer and let *time P(n)* be a poly(n) upper bounded number of deterministic computational steps (where time P(n) is not previously fixed for all possible programs **S**, but it is fixed for every one). **S** is a *restricted type X program* if and only if the following three conditions are satisfied:

1. **S** allows as input any **n**-bit word (member of arbitrary length **n** from $\{0, 1\}^+$).

2. The **S** behavior must be for each input one of the following:

i. **S** returns **0**;
ii. **S** returns **1**; or
iii. **S** does not halt (never returns any value).

3. The total **S** behavior must be for each **n** one of the following:

i. **S** returns in time P(n) **0** for all the $2^n$ possible inputs of length **n**; or
ii. **S** returns in time P(n) **1** for at least one possible input of length **n**.

**Note 1**: The presence of **S** is not to be decided – see Section 3.3.1. Testing whether a computer program is a restricted type X program will not be necessary to the proof. **S** will be given as an absolute assumption: It IS a restricted type X program, and this fact will NOT be under consideration: This is not a contradiction, definitely, since we can easily create innumerous programs of this type and – without need deciding about their types – produce a myriad of instances of the XG-SAT problem with them – see Sections 5.1 and 5.2, for details.

**Note 2**: There is no need that the polynomial running times involved in a proof must be previously fixed in order to be defined: For example, what is the fixed polynomial that upper bounds the running time of the reducer concerned in the Cook-Levin Theorem? There is no such fixed polynomial, since this running time depends on the NP problem whose instance is to be reduced to a Boolean formula, but the running time of this reducer is (and must be) polynomial, it is not undefined, of course, otherwise there would be no NP-Completeness. (This insight is formalized in the Def. 3.7.) Notice that it does not matter at all that we have a different time bound for each NP problem, but the same time bound for each instance of a fixed one, since for this reducer any instance from every NP problem is like just a mere *input* to a deterministic computer program: which is important here, in fact, is that that polynomial time bound is NOT *uniform*, whereas it is – without any contestation – considered very well defined. (Anyway, see [23] for a surprisingly claim about this issue.)

**Note 3**: The running time of a fixed program (or machine) **S** on those inputs for which it halts is bounded by a polynomial P(n) (which is a time-constructible function (for each fixed **S**), evidently [21]), hence there must be an equivalent machine (to each fixed **S**) which always halts, and still runs in deterministic polynomial time, of course. This, however, is not the main point: It is unimportant really whether there must be such an equivalent machine: What matters for my proof, after all, is that this equivalent machine (or program) cannot in general be constructed within deterministic polynomial time, at all, since that polynomial P(n) is *a priori* unknown or not given and – by Proposition 2.1 in [23] – it cannot be computed within deterministic polynomial time (see the detailed proofs in [23] and in Section 5).

**Note 4**: Into the traditional definitions of the classes P and NP, a polynomial P(n) must be fixed for whichever programs **S** (in order to the XG-SAT problem (Def. 3.1) is in traditional NP), and it is only over the class of all polynomial-time machines that such a polynomial is not fixed. However, into the new definitions of the classes P and NP (Defs. 3.5, 3.6 and 3.7), there is no need that there is a fixed polynomial P(n) for all possible **S** in order to the XG-SAT problem is in the new class NP (Def. 3.5) (see Proposition 3.1). Thus, the comparison with the Cook-Levin Theorem is herein quite well placed (in the note 2 above).

## 3. Definition of the XG-SAT Problem

**Definition 3.1.** Let **S** be a restricted type X program and let **n** be a finite positive integer. The *Extended General Problem of Satisfiability of a Boolean Formula* (*XG-SAT*) is the question "Does **S** return a value **1** for at least one input of length **n**?" Thus, in the XG-SAT question, the input is the pair ⟨**S**,$\mathbf{1^n}$⟩, clearly, where $\mathbf{1^n}$ is just **n** in unary form. Note that the specific and fixed time P(n) related to **S** is NOT given at all.

Be careful with a possible confusion made about the XG-SAT and the Bounded Halting problem (BH), defined over triples **w** = ⟨***M***,**x**,$\mathbf{1^k}$⟩, where ***M*** is a nondeterministic machine, **x** is a binary string, **k** is an integer, and **w** ∊ **BH** if and only if there exists a computation of ***M*** on input **x** that halts within **k** steps [12]: The XG-SAT is a very different problem, since the time P(n) is not given, and the program **S** into the pair ⟨**S**,$\mathbf{1^n}$⟩ *always* halts for at least one input of length **n**, but maybe **S** does not halt for all the other ones. Furthermore, the XG-SAT cannot be reduced in polynomial time to BH (– See Section 4.1). In order to understand why, verify that my XG-SAT problem is in the new [generalized] class NP (Def. 3.5), by Proposition 3.1, but it is not in that old traditional one.

## 3.1 Definition of well-formed string

**Definition 3.2.** Let **w** be a string from {**0**, **1**}$^+$. **w** is a *well-formed string* if and only if **w** has the form $\mathbf{1^+0s}$ – where $\mathbf{1^+}$ is a finite positive integer **n** encoded in unary form and **s** is the binary representation of the DTM (deterministic Turing Machine) that simulates a restricted type X program **S**. For **n** = 13, a well-formed string **w** would be, for instance, **111111111111101001000101101110010010101100100101011001001011110010010110...1**.

## 3.2 Definition of the XG-SAT as well-formed string acceptance testing to a language *L*

**Definition 3.3.** Let ***L*** be a formal language over the alphabet Σ = {**0**, **1**}. A well-formed string **w** ∊ ***L*** if and only if the DTM encoded into **w** returns **1** for at least one input of length **n**. The XG-SAT is the well-formed string acceptance testing to ***L***.

Note that as the size of a restricted type X program **S** is constant on **n** (|**S**|(n) = **c**), the length of the DTM that simulates **S** is constant too on **n** (|**s**|(n) = **k**), and then |**w**| = **n** + **1** + **k**. Thus, time P(n) is the same as time P(|**w**|) and time exp(n) is the same as time exp(|**w**|).

## 3.3 Class of the language *L* and Class of the $L_Z$-language *L*

***L*** is a nonrecursively enumerable (non-RE or non-Turing-recognizable) language [1], since it is undecidable whether or not an eventual result **1** from a computer program occurs within polynomial time [18], besides the undecidability even whether just it halts for some input [4].

(**Note**: The undecidability of the language ***L*** does NOT contradict the proof. The XG-SAT is not the undecidable decision problem **w** ∊? ***L***, but just the decidable one **well-formed string w** ∊? ***L***, as explained in Section 3.3.1, since a well-formed string **w** is *given* as an absolute assumption: **w** IS well-formed string, and this fact is NOT under consideration. See that exactly the same kind of statement holds to traditional formal languages, where the absolute assumption is that the strings to be tested are members from Σ* [1].)

<u>*Language Incompleteness*</u> – The computer theorists generally make a big mistake on definition of *computational decision problem*: They think that ones is the same thing that *languages*, as if all decision problems could be modeled as string acceptance testing to formal languages, like in [1, 5, 6]; however, there exist computational decision problems that can only be modeled as string acceptance testing to $L_Z$-*languages* (as defined in Section 3.3.1), not to languages, like the XG-SAT. (See Def. 3.8.)

Thus, all computer theorists generally say '*problem*' to mean '*language*' and vice versa. See below an excerpt of text of a preeminent Professor in the area, in [10]:

"*By Savage's theorem, any PROBLEM in P has a polynomial size family of circuits. Thus, to show that a PROBLEM is outside of P it would suffice to show that its circuit complexity is superpolynomial.*" [The words *PROBLEM* are lowercased in the original]

However, the set of all languages is a mere proper subset of the stronger and more powerful set of all $L_Z$-*languages* (all the computational decision problems), as established below.

### 3.3.1 More general definitions for NP and P and definition for $L_Z$-language

**Definition 3.4.** Let $L_Z$ be a language over a finite alphabet, $\Sigma$, and let $L \subseteq L_Z$. We will call $L$ an $L_Z$-*language*. If $L_Z = \Sigma^*$, then $L$ is a $\Sigma^*$-language, a *trivial* $L_Z$-*language*, which is the same as language ($\Sigma^*$-language = language). The complement of an $L_Z$-language $A$ is another $L_Z$-language $\bar{A} = L_Z - A$. Thus, $L_Z$-*language* is simply a generalization to *language* and a string acceptance testing to $L$ is a *computational decision problem* where the string to be tested is *necessarily* member from $L_Z$. If a *language* can be characterized as a *set*, an $L_Z$-*language* can be characterized as a *subset*, that is to say a *set into another*.

Observe that a string acceptance testing to $L$ is a *computational decision problem*, but $L$, rigorously, is not only a *language*, because $L \subseteq L_Z$, which is more restrict than simply $L \subseteq \Sigma^*$, which should hold if $L$ was only a language [6]. Thus, all the computational decision problems can be modeled as string acceptance testing to $L_Z$-languages, for to accept a string from any determined subset of $\Sigma^*$ is much more general than do it just from $\Sigma^*$, of course.

The main point herein is that the central relevance of the languages is originated in the fact that they model problems, not the inverse. Hence, great part of the Theory of Computation is about languages because of the mistake referred to in Section 3.3. When this mistake – that it is said as *mistake* because it leaves legitimate problems out of that old traditional definition – is fixed, the Theory of Computation will certainly study the generalization to language: The richer and stronger concept of $L_Z$-language.

A language over $\Sigma$ is a subset of $\Sigma^*$, and an $L_Z$-language is a subset of the language $L_Z$ over $\Sigma$. However, as $L \subseteq L_Z$ and $L_Z \subseteq \Sigma^*$, then $L \subseteq \Sigma^*$, which implies that all $L_Z$-languages are $\Sigma^*$-languages, or simply languages, too, naturally. Any language $L$ is also an $L$-language, and any $L_Z$-language $L$ is also a language $L$. In fact, if $L_Y \supseteq L_Z$ then any $L_Z$-language $L$ is an $L_Y$-language $L$, too. But the great advantage of the $L_Z$-languages is that string acceptance testing to ones can be much easier than to languages, because the strings $\mathbf{x}$ to be tested are in special form: $\mathbf{x} \in L_Z$ (this is an absolute assumption). Hence, if we know that all the strings to be tested are from a fixed language $L_Z$, then it is worth to model this problem as an $L_Z$-language; but if we do not know it, then we must model it as a simple language, of course.

Consequently, the concept of $L_Z$-language allows the insertion of previous knowledge about the form of the strings to be tested – when they were already constructed in special form or previously accepted by another machine – into traditional concept of language.

(**Note**: If the machine $M$ that decides an $L_Z$-*language* $L$ is fed a string **x** that is in $L_Z$, then $M$ *must* decide whether or not **x** is in $L$, anyway returning correct answer to **x** $\in$? $L$; on the other hand, if $M$ is fed any string that is not in $L_Z$, it may do whatever, returning anything, even *incorrect* answer to **x** $\in$? $L$ [$\Sigma$*-*language* $L$, in this case], or even not halting at all.)

For instance, the language $\{0^n1^n \mid n > 0\}$ over $\{0, 1\}$ is not regular, but verify that if $L_Z = \{0^n1^n \mid n > 0\} \cup \{1^n0^n \mid n > 0\}$, for example, then the $L_Z$-language $L_1 = \{0^n1^n \mid n > 0\}$ is regular and can be decided by the NFA $M = (\{q_0, q_1, q_2\}, \{0, 1\}, \delta, q_0, \{q_2\})$, where $\delta(q_0, 0) = \{q_2\}$, $\delta(q_0, 1) = \{q_1\}$, $\delta(q_1, 0) = \varnothing$, $\delta(q_1, 1) = \varnothing$, $\delta(q_2, 0) = \{q_2\}$, $\delta(q_2, 1) = \{q_2\}$, and there are not ε-moves.

Verify that this NFA $M$ recognizes the language $L = 0\{0, 1\}^*$ and $\{0^n1^n \mid n > 0\} = 0\{0, 1\}^* \cap (\{0^n1^n \mid n > 0\} \cup \{1^n0^n \mid n > 0\})$. In fact, this is not coincidence:

**Theorem 3.1.** If a machine $M$ (DFA, NFA, PDA, DTM, NTM, etc.) recognizes a language $L$, then $M$ recognizes any $L_Z$-language $L_1 = L \cap L_Z$.

*Proof*. Suppose that a string **x** $\in L_Z$-language $L_1$ was accepted by a machine $M$: Then, **x** $\in L_Z$ (this is an absolute assumption: All the strings to be tested must be member from $L_Z$) and **x** $\in L$ (the language that $M$ recognizes, regardless of the special form of **x**), which implies that **x** $\in L \cap L_Z$; on the other hand, if **x** $\in L \cap L_Z$, then **x** will be accepted by $M$, because **x** $\in L_Z$ (**x** can be tested) and **x** $\in L$ (**x** will be accepted by definition of string acceptance testing to languages), which implies that the $L_Z$-language recognized by $M$ $L_1 = L \cap L_Z$. □

See, thus, the proposed fix and generalization to the traditional formal definition for the class **NP** (Nondeterministic Polynomial Time) [14]:

**Definition 3.5.** Let $L$ be an $L_Z$-language. $L \in$ **NP** if and only if there is a binary relation $R \subseteq L_Z \times \Sigma^*$ and a known and given finite fixed positive integer $\boldsymbol{p}$ such that the following two conditions are satisfied:

1. For all $x \in L_Z$, $x \in L \Leftrightarrow \exists y \in \Sigma^*$ such that $(x, y) \in R$ and $|y| \in O(|x|^p)$; and
2. The language $L_r = \{x\#y : (x, y) \in R\}$ over $\Sigma \cup \{\#\}$ is decidable by a poly-time DTM.

A DTM that decides $L_r$ is called a *verifier* for $L$ and a **y** such that $(x, y) \in R$ is called a *certificate of membership* or *witness* of **x** in $L$. Note that – as **x** $\in L_Z$ (this is an absolute assumption, by Def. 3.4) – we do not need to describe what language $L_Z$ is allowed here. Hence, condition 2 does NOT require any knowledge about how to decide $L_Z$ in order to decide whether x#y is in $L_r$, plainly.

Verify that when $L_Z = \Sigma^*$ and the traditional definition for poly-time DTM there is utilized, the formal definition above is equivalent to the traditional one for the class **NP** – a set of mere languages –, which implies that this one is just a particular case of the proposed fixed definition. Consequently, we can name the traditional class **NP** as class **NP-SAT** (or, shortly, **SNP** or **NP**$_t$), where the Cook-Levin Theorem (with the hidden assumption referred to in [23]) and all the other mathematical truths on the traditional class **NP** continue holding in (replacing "$P \neq NP$" and "$P = NP$" by "*SAT is not in P*" and "*SAT is in P*", respectively, etc.).

Alternatively, we could call the true class **NP** defined above – an actual set of computational decision problems, or $L_Z$-languages – as class **NP-XG-SAT** (or, shortly, **XNP**), for example, but this naming method would be a mistake: A subset would have the name of the set and the set would have a derived name of the subset, which is hard to explain, confuse and damages the clearness of the notation. The same happens with the class **P** in Def. 3.6.

**Proposition 3.1.** XG-SAT is in class NP.

*Proof.* Into the Def. 3.5, for $\boldsymbol{L}$ modeling the XG-SAT, $\Sigma = \{\mathbf{0}, \mathbf{1}\}$, $\boldsymbol{L_z}$ is the set of all well-formed strings (as defined in Section 3.1), $\mathbf{p} = \mathbf{1}$, **y** is an **n**-bit word and $(x, y) \in R$ iff the program **S** encoded into the well-formed string **x** returns within polynomial time (poly(n), hence poly(|x#y|)) **1** for **y** as its input (a poly-time DTM that on ⟨**x#y**⟩ decodes and simulates **S** running having **y** as input is, in fact, the apt *verifier* for $\boldsymbol{L}$). Hence, $\boldsymbol{L}$ (XG-SAT) is in NP. □

Note that although the deterministic polynomial time $T(n) = O(n^{\mathbf{i}})$ that the witness predicate is decided is a different polynomial for each input **x**, XG-SAT is a single problem (it is false that any recursive decision problem is poly-time reducible to it, since T(n) is not previously fixed for all **S**, but it is fixed for every one, by Def. 2.1 – See the note 1 below, for details), where **i** does not depend on **n**, even though it does on **x**. Consequently, $(x, y) \in ?\ R$ is really decidable in deterministic polynomial time, by Def. 3.7, and the proof above is wholly correct: The XG-SAT is in NP, undoubtedly.

See that XG-SAT has strings of the form $\mathbf{1^n0s}$, where **s** is a DTM simulating a restricted X program **S** that accepts within polynomial time some string of length **n** (returning **1** for some **n**-bit input). Notice that we do NOT need to check whether **S** is a restricted X program, by Def. 3.4.

(**Note 1**: Suppose that someone says that the XG-SAT is not in NP, since its complexity class is really undefined, and it can be, for example, EEXP-Hard (for double exponential time), reasoning as below:

"Let $L$ be an EEXP problem, $M$ be the deterministic Turing Machine that solves $L$ in time $t(n) = 2^{2^{\wedge}\{poly(n)\}}$. Then we can reduce $L$ to XG-SAT as follows: Given an input **x** for the problem $L$, we construct a program **S** that ignores its input and simulates $M$ on input **x**. The promise is satisfied by the constant polynomial p(n') = t(|x|), and clearly (**S**,1) is an instance of XG-SAT if and only if $M$ accepts **x**."

Fortunately, constructions like above cannot disprove that XG-SAT is in NP, since they do not take into account that time P(n) is not previously fixed for all possible programs **S**, but it is fixed for every one, as stated in Def. 2.1 – hence, as $2^{2^{\wedge}\{poly(|x|)\}}$ is not upper bounded by any fixed poly(n), that program **S** is not a restricted type X program, and clearly (**S**,1) is NOT an instance of the XG-SAT.

Finally, see also that the function $2^{2^{\wedge}\{poly(|x|)\}} = t(|x|)$ is not constant, but depends on |x|. However, if **x** is fixed into that TM $M$ simulated by **S**, then this function is a constant (and then $M$ halts on **x** within only $O(1)$ steps, since $M$ and **x** are fixed independent of **n**); nonetheless, in this case, $M$ does not solve $L$, of course, and then the disproof above fails.)

(**Note 2**: Suppose, yet, that anyone else says that the XG-SAT is not in NP, since Proposition 3.1 is wrong, as long as either no poly-time TM can simulate a universal TM, or it – about the *verifier* for $\boldsymbol{L}$ that on ⟨**x#y**⟩ simulates **S** running having **y** as input – does not consider the running time of this simulation, which could be non-polynomial.

Fortunately yet, these refutations of Proposition 3.1 are equivocated, since a program **S** is always restricted (hence, it is NOT a universal TM), and the running time of the simulation of the program **S** (encoded into **x**) running having **y** (a *witness* of **x** in $L$) as input IS necessarily (must be) polynomial, since *time P(n)* is a time-constructible function (for each fixed **S**), by Def. 2.1.

See, however, this interesting review:

"– The author proposes that XG-SAT is in (promise-)NP but not in (promise-)P. He is right about the second part, but incorrect about the first part: XG-SAT is unconditionally not in promise-NP. He gives a simple but fallacious argument that XG-SAT is in promise-NP on p. 8. In note 2 on p. 8 he anticipates but rejects a counterargument, but he is wrong and this counterargument is essentially correct.

The reason is as follows: for any Turing machine *M* and positive integer *t*, we can form a machine $M_t$ that outputs **0** on all inputs except those of length *t*, on which it behaves like *M*. If *M* always halts and *M*'s behavior depends solely on its input length (call this latter restriction *semi-blindness*), then $M_t$ is always a restricted type-X machine.

It is known there exists a unary language *L* that is decidable, yet it is not in EXPTIME, hence not in NP. There is a *semi-blind* machine *M* that decides *L* correctly on each input having the form *1^t*. But if XG-SAT were in promise-NP, then we could solve *L* in NP: given input *x* of form *x = 1^t*, we decide whether *x* is in *L* by running the presumed NP verifier on the input ($M_t$, *1^t*), which obeys the promise. (If *x* is not of form *1^t*, then we can reject *x*.)"

Verify that that conclusion is not true: In order to try to decide that language *L* in NP, as proposed above, we must run the NP verifier on the inputs of form (*M, 1^t*), not ($M_t$, *1^t*), since to solve whether *M* accepts *1^t* is quite different from do it about $M_t$, for *M* is not the same thing neither has the same running time complexity as all the machines $M_1$, $M_2$, $M_3$, ... taken into account as a [countably infinite] set. The running time of all those $M_t$ is only $O(1)$, since *t* is a fixed constant into $M_t$, independent of **n** (|input|), while *L(M)* is not even in EXPTIME (hence, *M* is not a restricted type-X machine, at all), which implies, fortunately, that the input (*M, 1^t*) does not obey the promise in Def. 2.1, and then *L* cannot be decided in NP as proposed by that smart reviewer, and then the disproof above fails too.

See, also, another interesting and similar review:

"– Let *L* be any computable language, encoded in unary, and *M* a deterministic TM that solves *L*. The program $S=S_x$ takes its input **y**, and compare its length to **x**. If |y| = |x|, then S(y) simulates *M* on input **x**, and, if *M*(x) accepts, S(y) accepts. Otherwise, S(y) rejects. If |y| is any other value, S(y) rejects.

Clearly, this S runs in linear time, since all it has to do is count the length of y, **except** when |y| = |x|, but this is only finitely many exceptions, and hence doesn't change the asymptotic running time of S. To reduce *L* to XG-SAT: map **x** to the pair ($S_x$, 1^{|x|})."

Verify that that conclusion is not true too: By means of the same reasoning above, we could prove that that language *L* would be in NP, since $S_x$ and its input may be reduced to a Boolean expression in deterministic polynomial time (for the running time of $S_x$ is really only a fixed constant), and then this contradiction shows that this other disproof fails too.)

Note that even the language $L_r$ in item 2 above is, in fact, an $L_z$-language, where $L_z$ is the set of all strings of the form x#y.

In fact, after all, all complexity classes can be generalized with the concept of $L_Z$-*language*, like this new definition proposed for the class **P**:

**Definition 3.6.** Let $L$ be an $L_Z$-language. $L \in$ **P** if and only if for all $x \in L_Z$, $x \in? L$ is decidable by a poly-time DTM. Be careful with the traps: For example, all **$L_Z$-languages** $L_Z$ are trivially in **P** (where $L_Z$ can be *any* language, even non-Turing-recognizable ones), which does NOT mean that all **languages** $L_Z$ ($\Sigma^*$-languages) are in **P**, noticeably.

Notice that the proper definition of *deterministic polynomial-time computation* is more general here, without losing its more important characteristic: To be understood loosely as "feasible in practice", where the critique in [28] is not applicable:

**Definition 3.7: Poly-time DTM.** A DTM is said to be polynomial-time if its running time $T(n) = O(n^k)$, where $k = O(1)$, even that **k** depends some way on input. (n = |input|.)

Into the old traditional definition, **k** <u>must be</u> a fixed constant (which is in $O(1)$, obviously), but this stronger restriction is not essential to the vital matter: To maintain the character of vaguely practicable for deterministic polynomial-time computations. In XG-SAT, the T(n) of its *verifier* is in $O(n^k)$, where **k** depends on the **S** encoded into **w**, but it is in $O(1)$, since even that **k** cannot be computed [1] neither is given, it is a fixed constant for each fixed **S**, by Def. 2.1. Furthermore, the traditional definition of poly-time DTM asserts a hidden assumption: **k** <u>must be</u> *a priori* a <u>*known*</u> and <u>*given*</u> fixed constant, as revealed in [23].

See that if $T(n) = O(2^{poly(n)})$, for example, then $T(n) = O(n^k)$, where **k** ($poly(n) \log_n 2$) is not in $O(1)$, evidently, and is upper unbounded (for non-constant poly(n), of course): hence, in this case T(n) is not polynomial at all. The same happens with $T(n) = O(n^{\log n})$. If $T(n) = O(n^k)$, where **k** is, for example, the [arbitrary] position of the first 1 in **w** (or 1, if $\mathbf{w} = 0^n$), then **k** is not in $O(1)$ too, for those possible positions can be from 1 to $|\mathbf{w}| = \mathbf{n}$, hence in the extreme case $T(n) = O(n^n)$. On the other hand, if $T(n) = O(n^{g(n)})$, but now g(n) is upper bounded by a finite positive constant **k**, that is $\lim_{n\to\infty} g(n) < k$, then $T(n) = O(n^k)$, whence it is polynomial.

Some experts, as in [15], are asserting: "– The XG-SAT is not in NP (in the author's terms): The polynomial $n^k$ CANNOT depend on the input." However, this assertion is false, being true only for the old traditional definition of polynomial-time DTM, since in the new definition (Def. 3.7), the polynomial CAN definitely depend on the input – as long as that **k** is in $O(1)$. Think: This is just a matter of Math object definition, not of mathematical error or correctness, at all. We are not obligated to follow obsolete definitions only because they are established, unless the Science is finished (or dead). See Section 9.

Very important: Verify that these new definitions of the classes P and NP are simply good <u>generalizations</u> of the old traditional ones: Any traditional P or NP problem IS too, respectively, in the new class P or NP defined above (even though the converse is in general false, since these new generalized classes are strictly larger than the traditional ones), and any superpolynomial deterministic or nondeterministic problem is NOT in the new class P or NP, respectively, which proves that these generalizations are consistent and smooth.

### 3.3.2 A new NP genealogy

In fact, the traditional class NP (that we call herein $NP_t$) can be divided into two new disjoint classes: $NP_g$ (when that *p(n)* is known and given) and $NP_u$ (when *p(n)* is unknown or not given), where $NP_t = NP_g \cup NP_u$ and $NP_k \cap NP_u = \varnothing$. Into traditional beliefs, $NP_t$ is considerate equal to $NP_g$, and $NP_u$ is considerate equal to $\varnothing$, but these considerations take not account that the class $NP_u$ can be a genuine, useful and very important complexity class into the development of the Computational Complexity Theory, with great powerful applications in mathematically proven unbreakable in polynomial time public-key cryptography, for instance. By the way, see [27].

Into more formal terms, lets see the definitions for the two new disjoint classes that build the traditional class $\mathbf{NP_t}$: $\mathbf{NP_g}$ and $\mathbf{NP_u}$ (traditional Nondeterministic Polynomial Time when the involved polynomial time is or not given, respectively):

**Definition 3.8. $\mathbf{NP_g}$.** Let $L$ be a language over $\Sigma$. $L \in \mathbf{NP_g}$ if and only if there is a binary relation $R \subseteq \Sigma^* \times \Sigma^*$ and a known and given finite fixed positive integer $\boldsymbol{p}$ such that the following two conditions are satisfied:

1. For all $x \in \Sigma^*$, $x \in L \Leftrightarrow \exists y \in \Sigma^*$ such that $(x, y) \in R$ and $|y| \in O(|x|^p)$; and
2. The language $L_r = \{x\#y : (x, y) \in R\}$ over $\Sigma \cup \{\#\}$ is decidable by a polynomial-time DTM whose polynomial is fixed, known and given.

**Definition 3.9. $\mathbf{NP_u}$.** Let $L$ be a language over $\Sigma$. $L \in \mathbf{NP_u}$ if and only if there is a binary relation $R \subseteq \Sigma^* \times \Sigma^*$ and a known and given finite fixed positive integer $\boldsymbol{p}$ such that the following two conditions are satisfied:

1. For all $x \in \Sigma^*$, $x \in L \Leftrightarrow \exists y \in \Sigma^*$ such that $(x, y) \in R$ and $|y| \in O(|x|^p)$; and
2. The language $L_r = \{x\#y : (x, y) \in R\}$ over $\Sigma \cup \{\#\}$ is decidable by a polynomial-time DTM whose polynomial is fixed, but unknown or not given.

**Definition 3.10. $\mathbf{NP_t}$.** $NP_t = NP_g \cup NP_u$.

Let's see now the definitions for the class $\mathbf{NP_n}$: Non-uniform Nondeterministic Polynomial Time, as $NP_t$ but when the involved polynomial time is NOT fixed:

**Definition 3.11. $\mathbf{NP_n}$.** Let $L$ be a language over $\Sigma$. $L \in \mathbf{NP_n}$ if and only if there is a binary relation $R \subseteq \Sigma^* \times \Sigma^*$ and a known and given finite fixed positive integer $\boldsymbol{p}$ such that the following two conditions are satisfied:

1. For all $x \in \Sigma^*$, $x \in L \Leftrightarrow \exists y \in \Sigma^*$ such that $(x, y) \in R$ and $|y| \in O(|x|^p)$; and
2. The language $L_r = \{x\#y : (x, y) \in R\}$ over $\Sigma \cup \{\#\}$ is decidable by a poly-time DTM, as defined in Def. 3.7.

Now, the old traditional class NP ($NP_t$) is clearly seen simply as a proper subset of our new and legitimate extended class NP: $(NP_t \cup NP_n) \subset NP$ (as defined in Def. 3.5).

As always, in all the definitions above a DTM that decides $L_r$ is called a *verifier* for $L$ and a **y** such that $(x, y) \in R$ is called a *certificate of membership* or *witness* of **x** in $L$.

### 3.3.3 $L_z$-languages and Promise Problems

An *$L_z$-language $L$* can be considered as a *promise problem* $\prod$, as introduced by Alan L. Selman [Information and Computation, Vol. 78, Issue 2, (1988), pp. 87-98] and defined in [9], where the *promise* $(\prod_{YES} \cup \prod_{NO}) = L_z$, $\prod_{YES} = L$, $\prod_{NO} = L_z - L$, and its restricted alphabet {0, 1} is generalized to any finite alphabet Σ. Nonetheless, notice that the concepts, notation, generality, power and applicability of the *$L_Z$-languages* are clearer, richer, simpler, conciser, more elegant, aesthetic and stronger than ones of the *promise problems*.

## 3.4 More general definition for Computational Decision Problem

Note yet that the definition of *computational decision problem* used herein is also more general, without losing its more essential attribute: To model *all* real computer-based questions – not only a small part of them – having one and only one answer from two alternatives [16]:

**Definition 3.8:** A *computational decision problem* is any arbitrary **Yes**-or-**No** (**True**-or-**False**) question on a finite or countably infinite set of inputs (strings of any finite length over a finite alphabet Σ), where these ones are necessarily member from another determined set (or consistently the set of inputs *of obligatory specified form* for which the problem returns **Yes** (**True**)). Equivalently, decision problems are completely isomorphic to $L_Z$-languages of strings, and can always be modeled as string acceptance testing to $L_Z$-languages.

Into the traditional definition for computational decision problem [1, 16], using plain languages, the inputs for a problem are simply from Σ*, whereas for this more general definition they are from any arbitrary subset of Σ*. So, we can consider a *traditional* problem (*language*) as a *set*, and a *more general* one (*$L_Z$-language*) as a *subset* of a *set*. Hence, the set of all languages ($L_Z$ = Σ*) is just a little proper subset of the set of all $L_Z$-languages ($L_Z$ = any subset of Σ*). So, only *one* set characterizes language, but we need *two* sets for $L_Z$-language.

See yet that, for this generalization, the strings to be tested (into a string acceptance testing to an $L_Z$-language) are necessarily (must be) member from $L_Z$ (whatever $L_Z$ is), where this fact IS an absolute assumption and IS NOT under consideration. Verify that exactly the same kind of statement holds to traditional formal languages, where the absolute assumption is that the strings to be tested (into a string acceptance testing to a language) are always necessarily (must be) members from Σ*.

### 3.4.1 The falseness of the Cook-Levin Theorem

**Theorem 3.2.** The Cook-Levin Theorem (CLT) is false.

*Proof.* See it in [23]. There are some comments on it in [24].

#### 3.4.1.1 How can a Theorem be false?

Since a *theorem* is an absolute mathematical truth, how can the Cook-Levin Theorem be false?

See it in [23].

## 4. Old demonstration that the XG-SAT is in NP

Given **n** and a restricted type X program **S**, the question "Does **S** return a value **1** for at least one input of length **n**?" can be decided in nondeterministic poly-time (time NP(n): as time P(n), using nondeterminism), since can be constructed a universal NTM (nondeterministic TM) that simply simulates the running of **S** and tests it for all $2^n$ possible inputs of length **n** at the same time ("on parallel") and verifies in time NP(n) the returns: If they are **0** for all the inputs, then the NTM will answer "No" after the conclusion of the last computation path (branch); on the other hand, if at least one return is **1**, then the NTM will answer "Yes" at the end of the first path that returns **1**, regardless of whether the others are yet running. One and only one of these two events must happen in time NP(n), by Def. 2.1.

## 4.1 The SAT as particular case of the XG-SAT

Any Boolean formula **E** with **n** variables can be simulated in polynomial time by a simple restricted type X program **S** (where each bit from input represents one variable) where the returns from **S** for all $2^n$ possible inputs of length **n** represent the results from **E** for all assignment of truth values to the variables (**0** represents **False** and **1** represents **True**). For instance, if **E** has two variables and **E**(F,F) = **F**, **E**(F,T) = **F**, **E**(T,F) = **T**, **E**(T,T) = **F** then **S**(00) returns **0**, **S**(01) returns **0**, **S**(10) returns **1**, **S**(11) returns **0**; finally, **S** returns **0** for all inputs of length different from **n**.

Let **S** be the restricted type X program below, an example of such a simulator:

```
01. S(string input) // input is an n-bit word where each bit from it represents one variable from
                    // a Boolean formula E with n variables.
02. {
03.   string E := "x1 ∧ ((¬x1 ∧ x10) ∨ ¬(x1 ∧ ¬(x10 ∨ x11) ∧ ¬x1))"; // any Boolean
formula of length m can be placed herein and the number of variables in it can be determined
in time P(m), naturally, where m = O(n log n) – which implies that time P(n) is the same as
time P(m).
04.   return(Satisfier(E, input)); // Satisfier is a simple function that returns 1 if the
assignment of the input's truth values to the variables of E, as above, satisfies it; otherwise it
returns 0. If the length of input is different from the number of variables in the E, then it
returns 0, too. Satisfier can (and must) run in time P(n), of course, and must not use short
circuit logic in order to its running time be the same for any input of length n.
05. }
```

Thus, if there exists a Poly(n)-time decider for XG-SAT, then it can decide the SAT in polynomial time (time P(n)) too: It's enough to place the Boolean formula **E** to be tested into the code **S** above, to construct, setting **n** as the number of variables in **E**, an instance **w** of the XG-SAT with **S**, and to decide in time P(n) whether **w** is in the language ***L*** defined in Section 3.2, which is the same as **E** to be satisfiable, naturally. See that, in fact, this process constructs a different restricted type X program for each different **E** to be tested.

On the other hand, if a restricted type X program **S** returns in time P(n) **0** or **1** for all $2^n$ possible inputs of length **n** (where its running time is the same for any one), and **0** for all inputs of length different from **n**, then **S** can be considered a simple simulation of a Boolean formula with **n** variables, as above, which implies that the **SAT** is just a particular case of the **XG-SAT**.

Moreover, given an instance of the XG-SAT constructed with a program **S** that returns in time P(n) **0** or **1** for all $2^n$ possible inputs of length **n** and a running time (upper bounded by a known and given poly(n), naturally) that bounds the entire running time of the NTM that decides in time NP(n) this instance, this NTM can be reduced to a Boolean formula that represents its entire processing, by means of the Cook-Levin Theorem (Cook's Theorem) [1].

Verify that the running time of this NTM is equal to a fixed value more the running time of running the simulation of **S**, that can be determined by way of running the simulation of **S** for some input of length **n** and counting the running time in order to it returns the result **0** or **1**, because if it is simulating a Boolean formula with **n** variables, then its running time is the same for any input of length **n**. Notice that if **S** is not simulating a Boolean formula, then this way of determining its running time for some input – and then the running time of the NTM that solves the instance of the XG-SAT constructed with this **S** – does not work in general, for **S**, in this case, can never halt for some inputs, or its running time can be different for different inputs, even though ones of same length.

*NP-Completeness Incompleteness* – Thus, as a restricted type X program does not necessarily halt for all the $2^n$ possible inputs of length **n** and the running time of the universal NTM that decides in time NP(n) any instance of the XG-SAT cannot be upper bounded by any fixed poly(n) – by Def. 2.1 –, the Cook-Levin Theorem cannot in general be used on an instance of the XG-SAT to reduce it within polynomial time to an instance of the SAT. In fact, the SAT is very easy compared to the XG-SAT, and the speed of the nondeterministic computation is much greater than we have believed it. (See Section 3.4 and [23].)

Notice that, despite the time P(n) cannot be upper bounded by any fixed poly(n), it is neither vague, undefined, nor undetermined: For each restricted type X program **S**, there is exactly one specific and fixed polynomial in **n** that bounds the necessary number of deterministic computational steps in order to **S** returns **0** for all inputs of length **n**, or **1** for at least one, where **n** is an arbitrary positive integer, but this polynomial in **n** is fixed, by Def. 2.1, even though it is not given at all.

The main astonishing idea that was able to separate the classes P and NP: This time P(n) depends on **S**, but it does not depend on **n**, which makes all the difference and changes forever almost anything in the Computational Complexity Theory…

## 5. Demonstration that the XG-SAT is not in P

**Theorem 5.1.** P $\neq$ NP.

*Proof.* As demonstrated in Sections 3.3.1 and 4, any instance of the XG-SAT can be recognized in nondeterministic polynomial time. However, can it be recognized in deterministic polynomial time?

By hypothesis, consider that it can: In this case, must exist a DTM **Q** that – given a positive integer **n** and a restricted type X program **S** into **w** – answers correctly within polynomial time the question "Does **S** return a value **1** for at least one input of length **n**?" (If **w** is in XG-SAT, then **Q**(**w**) = "Yes", else **Q**(**w**) = "No"). **Note**: All the inputs for the program **S** in this Section are of length **n**.

**Proposition 5.1.** The DTM **Q** is, in fact, a real computer program. Although it may work entirely in a different way from someone would expect from the method that the XG-SAT was defined, **Q** cannot be a magical or dream machine, since it must be an actual machine.

So, let **W**: $\Sigma^* \times L_Z \rightarrow \mathbf{N}$ be a function with a DTM and a well-formed input for it as arguments, where if **W**(**Q**, **w**) = **m** (**Q** can simulate the running of **S** into **w** and test some inputs for **S** in such a simulation – considered herein a step-by-step process running **S** into **Q**), then **m** is the number of inputs for **S** simulated by **Q** in this process: $0 \leq \mathbf{m} \leq 2^n$.

**Note**: It does not matter for this proof whether **W** is a computable function or not; and if **X** is not a DTM or is not interested in the XG-SAT problem, then **W**(**X**, **w**) is defined as 0.

Thus, in order to answer the question, there are no miracles: **Q** can act into only four possible ways (where **m** = **W**(**Q**, **w**)):

1. **Q** simulates the running of **S** for:

   i. All the possible inputs (**m** = $2^n$);
   ii. All the inputs from an arbitrary nonempty proper subset of all them (0 < **m** < $2^n$); or
   iii. Only one input (or all from a nonempty proper subset of all them) previously computed whose return decides the question (**m** = **d** < $2^n$).

2. **Q** does not simulate the running of **S** at all (**m** = 0).

*Proof.* These ways are exhaustive: Either **Q** simulates the running of **S** or not; and, if **Q** simulates the running of **S**, then it can test on it all the possible inputs (1.i); arbitrarily less than all ones (1.ii); or just one (or all from a nonempty proper subset of all them) that was anyway previously computed whose return decides the question (1.iii). Unfortunately, there are no more alternatives besides that ones. (Note: Into ways (1.ii) and (1.iii), **m** must be polynomial in **n** in order to **Q** can decide the XG-SAT in deterministic polynomial time, of course.)

As well, the running time of a universal NTM that decides in time NP(n) the XG-SAT – as in Section 4 – cannot be upper bounded by any fixed poly(n). Moreover, a program **S** does not necessarily halt for all its possible inputs. Furthermore, the time P(n) in Def. 2.1 cannot be upper bounded by any fixed poly(n), too. Thus, in general, cannot exist any fixed poly(n) number of TM configurations that represents the entire processing of **S**.

Additionally, as to find the input whose return decides the question and simulate the running of **S** only for this input is impossible (see in Way 1.iii below), the particular fixed running time P(n) of a specific restricted type X program cannot be computed within any fixed poly(n) upper bounded number of deterministic computational steps. Hence, an instance of the XG-SAT cannot be reduced within polynomial time into another one of another poly-time decidable problem, because the reducer machine must run within polynomial time in this case, but it cannot previously know or compute what upper bounds that time P(n), by Proposition 2.1 in [23]. □

Suppose, however, that someone claims, with the following argument, that the P ≠ NP proof of mine fails:

"– The author assumes that the 4 ways mentioned are the only way to solve the problem. Why can't the DTM **Q** decide the question some other way?"

The answer is not complicated: **Q** cannot decide the question by some other way because there is no another possible way to decide the XG-SAT besides the four ones mentioned above: These ways do not specify type, structure, form, code, nature, shape or kind of computation, neither structure (or lack thereof) of data – but just the **number (m) and kind of inputs (arbitrary or computed) tested** in eventual simulated running of **S** –, into *any* running of *any* DTM that tries to decide the XG-SAT: (1) all inputs (**m** = $2^n$); (2) arbitrary ones less than all (0 < **m** < $2^n$); (3) computed ones less than all (**m** = **d** < $2^n$); or (4) none (there is no simulating **S** into **Q** at all) (**m** = 0).

Can there be some other way? No, by a reasoning similar to *pigeonholes* from *pigeonhole principle*: Either **Q** simulates **S** or not. And simulating **S** for more than all inputs – or for any subset with exponential number of them – leads to exp(n) running time, as explained in the Way 1.i; less than none go to negative number of inputs, which makes no sense in actual computations; and between these limits the number and kind of inputs for eventual simulated running of **S** must be one from the four mentioned above. Consequently, all the possible deterministic computations to decide the XG-SAT are really into one from these four ways.

Can we create new ways to decide in deterministic polynomial time the XG-SAT combining the four ones? Unfortunately, no way: The way 1.i is useless to decide in deterministic polynomial time the question; the way 1.ii is useless to decide in any time the XG-SAT; and the combination of the ways 1.iii and 2 results simply in the way 1.iii – when none result from the simulation is used by **Q** in order to answer the question, which is a case treated below in the way 1.iii.

Hence, claims like above do not go to refute this P ≠ NP proof.

Note, yet, that the method utilized in this proof cannot be adapted to decide whether SAT is in P, because if a program **S** is simulating a Boolean formula with **n** variables, it *must* always halt for all the possible inputs, and its running time *must* be the same for any input; however, these additional restricted conditions cannot be held in general restricted type X programs, like ones in the proofs of the Props. 5.2 and 5.3 below.

Hence, to decide whether an arbitrary general deterministic computer program computes determined output for determined input (which is undecidable, by the Rice's Theorem [11]) cannot be reduced to SAT as it does to XG-SAT (as demonstrated in these proofs), and then any attempt to adapt my proof to solve whether SAT is in P is condemned to fault.

See that if **S** is simulating a Boolean formula with **n** variables, then the Rice's Theorem cannot be applied to the **S** behavior for any input, since it is restricted for all the possible ones.

Finally, suppose that else one tries to refute the proof saying:

“This proof follows a common theme: Defines an NP problem with a certain structure, argues that any algorithm that solves that problem must work in a certain way and any algorithm that works that way must examine an exponential number of possibilities. But we can't assume anything about how an algorithm works. Algorithms can ignore the underlying semantic meaning of the input and focus on the syntactic part, the bits themselves.”

As in the previous “refutation” of my proof, the answer is also not too complicated: If the DTM **Q** ignores the underlying semantic meaning of **w** and focus on its syntactic part, the bits themselves, considering **w** just a series of bits, then this approach only places **Q** into the Way 2 – where **Q** does not simulate the running of **S** at all (**m** = 0) –, and then the proof continues to hold, naturally.

Shortly, the spirit of the proof is very simple: The XG-SAT is decidable by brute-force search because whether or not **S** returns **1** for at least one input from all the $2^n$ possible ones is decidable, whereas whether **S** returns **1** for at least one from a nonempty proper subset of them is in general undecidable (since **S** can even not halt for any input from such a proper subset), by Def. 2.1, which does that all the other ways to decide the XG-SAT (without brute-force searching) be absolutely hopeless.

Consequently, we can say that the profoundest question in Computational Complexity Theory was solved by this plain ingenious characterization, the Def. 2.1!

Be brave and see below that all these four exhaustive ways to decide in deterministic polynomial time the XG-SAT fail:

### Way 1.i Q simulates the running of S for all the possible inputs (m = $2^n$):

The obvious way to implement the DTM **Q** is to construct a universal DTM that simulates the running of **S** and submits to it each one of the $2^n$ possible inputs, verifying whether it returns **1** for at least one (in a breadth-first search, to avoid running forever in a computation path that does not halt): If all returns are **0**, then **w** is not in XG-SAT; otherwise, then it is. Exactly one of these two events must happen in time NP(n), by Def. 2.1.

Nevertheless, this brute-force method, on worst case, can decide the problem only at the end of testing all the $2^n$ possible inputs, in time exp(n).

### Way 1.ii Q simulates the running of S for all the inputs from an arbitrary nonempty proper subset of all them (0 < m < $2^n$):

Note that to simulate the running of **S** only for a polynomial number of arbitrary inputs (or just for a number of them less than all the possible ones – for instance: $\mathbf{n}^{\log n}$) does not work: Even the test of $2^n - 2$ inputs on the simulation cannot help to decide whether **S** returns **1** for some from the two not simulated ones (in fact, this simulation cannot help to decide even whether **S** simply halts for a specified input from these two ones).

Moreover, even the simple question whether **S** halts for at least one input from an arbitrary nonempty proper subset of the set of all the $2^n$ possible inputs is undecidable, of course, by Def. 2.1. (Obs.: This question is only decidable for the set of all the $2^n$ possible inputs: The answer is always "True" – **S** halts for at least one input –, by Def. 2.1.)

### Way 1.iii Q simulates the running of S only for an input (or inputs) previously computed (m = d < $2^n$):

**Proposition 5.2.** A DTM **Q** cannot compute, without simulating the running of **S** for all the $2^n$ possible inputs, a nonempty proper subset of ones, where the return from **S** for one of them decides the question, and then to simulate the running of **S** only for these inputs to decide the XG-SAT.

*Proof.* Let a well-formed string **f** be constructed with an arbitrary **n,** and let the restricted type X program **F** be below, where **Q** was, by the Turing-Church Thesis, translated into a computer program where it was included the instruction Simulated_by_Q[e] := True; just before any instruction of this program that starts the simulation of **F** for any input **e** (Simulated_by_Q is a global variable of type dynamic array or vector of Booleans values that was initialized with False in all its positions).

We call **Q'** to this program derived from **Q**. Verify that if **Q** runs in polynomial time, then **Q'** also do it, of course, and the behaviors and results from **Q'** and **Q** are the same.

```
01. F(string input) // F is a restricted type X program, since Q' and R are supposed poly-time
                    // DTMs, and F will either return only 0's or at least one 1
02. { n := length(input);
```

```
03.   if (R(input) = “Yes”) return(0); // R returns always within polynomial time “Maybe”
04.   if (R(input) = “No”)  return(1); // Thus, R does not matter to the behavior of F
                                       // But Q does not know it: See its work is very hard!
05.   concurrent_ifs // only one of the returns can close the concurrent instructions block
06.   {               // below, where the two if’s run concurrently
07.     { if (Simulated_by_Q[input])  return(0); } // There will be d of these returns …
08.     { if (Q’(f) = “Yes”)          return(0);  else return(1); } … and 2^n-d of these ones
09.   }
10. }
```

Now, as **d** < $2^n$, then **Q’** will unavoidable answer incorrectly “No”, after **F** returns **0** for all the **d** simulated inputs, since **F** will in this case return **1** for all the non-simulated inputs (there is at least one, since **d** < $2^n$), because for these ones there will be chance for the second **concurrent_if** (line 08) to detect at some moment the answer “No” from **Q’**, and then to return **1**. Note that **Q’** cannot answer “Yes”, because **F** will always return **0** for all the **d** simulated inputs, by the first **concurrent_if** (line 07), and the answer from **Q’** is based in the returns from **F** for all the simulated inputs. See that the contents of the global variable Simulated_by_Q (line 07) is known into **F**, because **Q’** is concurrently running (line 08).

On the other hand, if **d** = $2^n$, then it answers correctly “No”, after **F** returns **0** for all inputs, since **F**, in this case, returns **0** for all ones because the return of all **0** is by the first **concurrent_if**, when then there is no chance for the second **concurrent_if** to detect the answer “No” from **Q’**, and then to return **1**. Unfortunately, **Q** can, by this way, decide the XG-SAT just in time exp(n), as was treated in Way 1.i.

See that if **Q** does not simulate the running of **S** for any input at all (**d** = 0) – or if none result from the simulation is used by **Q** in order to answer the question –, then it will inevitably answer incorrectly at some moment, by diagonalization that exists into string **f**, on the second **concurrent_if** of **F** (line 08).

Finally, suppose that **Q** could decide whether there is diagonalization into string **f**. In this case, **Q** could stay running forever, without simulating the running of **F** (or simulating it and no using the results **0** from this simulation to answer “No”) and no returning anything at all, which would imply that **f** is not a well-formed string, and then **Q** would not be incorrect. Alternatively in this case, **Q** could attempt to decide the question either without simulating the running of **S** for any input at all (**d** = 0), or simulating it for some inputs and ignoring the results **0** from this simulation (there is no result **1**, of course), either considering thereby **w** just a bit string, or engaging in more indirect reasoning about the code of **S**, as in the Way 2 below.

However, **Q’** can in general be any arbitrary deterministic program and can compute using or not the value of ***input*** for **F**, besides its proper input (the string **f**). Here, if **Q’**(**f**) runs in polynomial time either returning always “Yes” for all values of ***input*** not simulated by **Q**, or returning another result for at least one, then **f** is a well-formed string (independently of the behavior of **Q**(**f**)), and there is no diagonalization into it. Consequently, there is diagonalization into string **f** if and only if **Q**(**f**) = **Q’**(**f**) independently of the value of ***input***.

Hence, if **Q** can decide whether there is diagonalization into string **f**, without simulating the running of **S** for all the $2^n$ possible inputs, then **Q** can decide whether the language of a given arbitrary TM (**Q’**) has a particular nontrivial property (**Q’** accepts **f** if and only if itself (**Q**) does it; in other words, **Q**(**f**) = **Q’**(**f**) independently of the value of ***input***). However, this problem is undecidable, by the Rice's Theorem (reflect: **Q’** could be *any* computer program). Hence, **Q** cannot decide it; thus, **Q** is condemned to fault, too: without knowing neither computing whether there is diagonalization into string **f**, to answer incorrectly the question, by the diagonalization above; or to simulate the running of **F** for all the $2^n$ possible inputs (**d** = $2^n$), in time exp(n). □

Note that, in general, the Rice's Theorem can be applied to the **S** behavior for a nonempty proper subset of all the $2^n$ possible inputs (because this behavior is arbitrary, by Def. 2.1: in this case, **S** can not halt for any input from this subset), but cannot do it to the total **S** behavior for all ones (since this behavior is restricted, by Def. 2.1: in this case, either all results from **S** are **0** or at least one is **1**).

## Way 2. Q does not simulate the running of S at all (m = 0):

If the running time of **Q** depends on the one of the restricted type X program **S** into **w** for some input (where if **S** does not halt for any input, then **Q** does not halt at all, too), which occurs when **Q** acts reducing **w** into instance of another problem or simulating the running of **S**, then the use of the diagonalization method in order to demonstrate that **Q** cannot decide the problem fails, since **Q** does not have to be as restricted as **S**. But, as these running times are independent ones in the special case treated here, where **w** is considered either just a bit string, or **Q** decides whether **S** returns **1** for some input by engaging in more indirect reasoning about the code of **S**, without simulating it at all, then we can use diagonalization in order to demonstrate that **Q** cannot decide the XG-SAT. E.g., if **Q** converts a problem in **E** without $2^{n/10}$-size circuits into a PRG which fools $n^c$-size ones, for any fixed **c**, then **Q** is here.

Note that if the running time of **Q** is anyway always greater than one of the restricted type X program **S** into **w** for some input (where, remember, if **S** does not halt for any input, then **Q** does not halt at all, too), then **Q** is, maybe indirectly, simulating the running of **S** for this same input or reducing the instance of the XG-SAT constructed with **S** to some instance of another problem, of course. In general, to reduce within polynomial time an instance of the XG-SAT is impossible, by the proof of the Proposition 5.1. We will see below that to decide the XG-SAT without simulating the running of **S** – or do it in a running time upper bounded by any fixed (or even non-fixed) integer polynomial function of |**w**| – is impossible, too:

**Proposition 5.3.** A DTM **Q** cannot, without simulating the running of **S** for any input, decide the XG-SAT problem.

*Proof.* Let **S** encoded into **w** be the program:

```
01. S(string input)
02. {
03.   n := length(input);
04.   if (integer(input) = 2^n – 1) { if (T(w) = “Yes”) return(0); else return(1); } else return(0);
05. }
```

Where **T** is an arbitrary deterministic program. Hence, if **Q** can, without simulating the running of **S** for any input, decide the XG-SAT problem, then it can decide whether the language of **T** has a particular nontrivial property: **T**(**w**) != **Q**(**w**), since otherwise then **Q** cannot answer anything within polynomial time, because whatever it answers will be incorrect, by diagonalization that holds in this case, in line 04 (**S** returns **0** for all inputs of the form different from $\mathbf{1^n}$, and the answer from **T**(**w**) is inverted and returned by **S** when it process the input of the form $\mathbf{1^n}$ – what is the same as ***integer(input) = 2^n – 1***: See that there is only one input of this form for each **n**. Thus, if **T**(**w**) returns ***“Yes”***, then **w** is not in XG-SAT, and vice versa). However, if **T**(**w**) != **Q**(**w**) and **T**(**w**) runs in polynomial time, then there is no diagonalization into **S**, and **Q** must decide what **T**(**w**) returns to decide the question. That is, before **Q** answers whatever within polynomial time, it must decide whether

**T**(**w**) = **Q**(**w**), to avoid that the diagonalization above forces it to error (reflect: **T** can be *any* computer program).

Nevertheless, this problem is undecidable, by the Rice's Theorem; hence, the DTM **Q** cannot, without simulating the running of **S** for any input, decide the XG-SAT. □

Observe that if **Q** tries to test whether the string **w** above is in XG-SAT (where **Q**(**w**) can return: "Yes"; "No"; or it does not halt) simulating the running of **S** for all the possible inputs, then **S** returns in time P(n) **0** for all ones of the form different from $\mathbf{1^n}$, but, if **T**(**w**) = **Q**(**w**), then the simulation of the running of **S** never halts and never returns any value for the input of the form $\mathbf{1^n}$, by infinite regress, which implies that **Q** does not halt and never answer incorrectly, even though without deciding whether **T**(**w**) = **Q**(**w**), obviously, since it remains forever waiting the return from this simulation for input of the form $\mathbf{1^n}$, deluded without knowing that it never does. Notice that if **T** answers anything within polynomial time, then **w** is a well-formed string; otherwise, then it is not.

Perceive that to state that **Q** answers whatsoever if and only if **w** is a well-formed string does not work, because, as demonstrated in Section 3.3, ***L*** (the set of all well-formed strings) is a non-RE language, which implies that **Q** cannot decide whether **w** is a well-formed string in order to decide accordingly whether it can answer anything without mistaking. That is, in order to **Q** works in this case, it must assume absolutely that **w** is a well-formed string, and then this assumption implies that it is really true, and that **Q** for the input **w** returns within polynomial time incorrect answer, by the diagonalization above into **S** (line 04).

**Proposition 5.4.** A DTM **Q** with running time upper bounded by $f(|\mathbf{w}|)$, where $f$ is a fixed [or even non-fixed] integer polynomial function of |**w**|, cannot decide the XG-SAT.

*Proof.* Let **S** encoded into **w** be the program:

```
01. S(string input)
02. {
03.   n := length(input);
04.   if (integer(input) = 2^n – 1) {
05.     concurrent_ifs {
06.       if (T(w) = "Yes") return(0); else return(1);
07.       if (Timer > |w|^k)   return(1);
08.     }
09.   }
10.   else return(0);
11. }
```

Where **T** is an arbitrary deterministic program, **k** is an arbitrary fixed finite positive integer, ***Timer*** counts (in another "thread" or "concurrent process") the running time of **S**, and the two internal ***if*** are evaluated concurrently while **T** runs for the input **w**. Thus, as the functions ***length*** and ***integer*** are poly(n)-time, **S** is a restricted type X program, regardless of the behavior of **T**. See that, as **S** returns **0** for all inputs of the form different from $\mathbf{1^n}$, if **T**(**w**) answers in running time less than $|\mathbf{w}|^{\mathbf{k}}$, then its answer is inverted when **S** process the input of the form $\mathbf{1^n}$, and **T** is forced to the error if it tries to be a decider for XG-SAT. Therefore, as **T**(**w**) can be equal to **Q**(**w**) – (if **Q** and **T** are equivalent, for instance) – even though **Q** cannot decide whether it is true that **T**(**w**) = **Q**(**w**), by the Rice's Theorem (reflect: **T** can be *any* computer program) –, this implies that, for large enough **k** (when $|\mathbf{w}|^{\mathbf{k}} > f(|\mathbf{w}|)$, **Q** fails: It cannot know or compute that it cannot in this case answer correctly the question in running time less than $|\mathbf{w}|^{\mathbf{k}}$, by the diagonalization above.

Hence, **Q** is again condemned to fault: To answer incorrectly the question before $|\mathbf{w}|^{\mathbf{k}}$ computational steps, for some large enough **k**. Note yet that **Q** cannot adjust $f(|\mathbf{w}|)$ in order to it is always greater than $|\mathbf{w}|^{\mathbf{k}}$, since this polynomial is *a priori* unknown or not given and – by Proposition 2.1 in [23] – it cannot be computed within deterministic polynomial time. □

Observe again that if the DTM **Q** decides the XG-SAT simulating the running of **S**, then **Q** cannot run in time upper bounded by any fixed polynomial function of |**w**| (in fact, none TM – DTM or NTM – that decides the XG-SAT can do it), undoubtedly, by Def. 2.1.

## Conclusion:

As demonstrated above, all the four exhaustive possible ways to decide in deterministic polynomial time the XG-SAT fail: Consequently, there exists a computational decision problem that can be decided in nondeterministic polynomial time, but not in deterministic polynomial time, which implies **P** ≠ **NP**, naturally, in our sad computational world. □

For this reason, by union of the Rice's Theorem, the diagonalization method and the complexity classes P and NP, this proof is a beautiful unification and an amazing synthesis between the Computability Theory and the Computational Complexity Theory.

Lastly, someone can say that if a fixed and known *p(n)* ≥ time P(n) of the program **S** into **w** is given (see this one is not deterministic poly-time computable, by Proposition 2.1 in [23]), then the instances of the XG-SAT can be reduced to Boolean formula ones by Cook-Levin Theorem, and then if the SAT is decidable in deterministic poly-time, then the XG-SAT is too. Big idea!

This conclusion is erroneous, however, since knowing a fixed polynomial *p(n)* ≥ that time P(n) is unnecessary to decide the problem (the universal NTM in Section 4 and the universal DTM in Way 1.i decide the XG-SAT without knowing this information (or without such an input), naturally), proving that nondeterministic computation is fundamentally much more faster than deterministic computation, and that the brute-force search is unfortunately unavoidable in the real-world computations (I'm very sorry): To verify a correct answer is definitely very easier than find it, naturally.

## 5.1 Running time of the functions into programs

About running in time P(n) and time greater than P(n), let the function be:

```
01. Poly_Function(string input)
02. {
03.   int i, counter := 0, n := length(input);
04.   for i := 1 to n^10 { counter := counter + 1; } // poly(n) upper bounded running time
05.   if (counter > 100) return(1); else return(0);
06. }
```

The function above evaluated at string input is just a number, naturally. But we can decide that its running time is poly(n) upper bounded, where **n** = |input|. We don't need a TM to decide it. On the other hand, let the function be:

```
01. SuperPoly_Function(string input)
02. {
03.   int i, counter := 0, n := length(input);
```

```
04.   for i := 1 to 2^n { counter := counter + 1; } // exp(n) upper bounded running time
05.   if (counter > 100) return(1); else return(0);
06. }
```

Of course, the running time of the function above is exponential in **n**. We know countless functions as the ones above [2] to use them in order to make restricted type X programs. Constructing restricted type X programs using algorithms with known running time is human work, not TM computation [2].

## 5.2 Example of construction of an instance of the XG-SAT

Let the restricted type X program **S** be:

```
01. S(string input)
02. {
03.   remainder := mod(integer(input), 2);  // remainder on division of input (converted into
                                            // integer) by 2
04.   if (remainder = 0) return(Fun2(input)); // returns the value returned by Fun2 and halts
05.   if (remainder = 1) return(Fun1(input)); // never halts
06. }

07. Fun1(string input) {
08.   do { input := "1"; } while (1 = 1); // infinite loop
09.   return(1);
10. }

11. Fun2(string input) {
12.   int i, counter := 0, n := length(input);
13.   for i := 1 to n^10 { counter := counter + 1; } // poly(n) upper bounded running time
14.   if (counter > 0) return(1); else return(0);
15. }
```

Thus, we can simply convert this program **S** into a DTM ***M***, translate it into a binary form **s**, and then construct the well-formed string **w** = **111111110s**, an instance of the XG-SAT. Here, constructing XG-SAT instances, it stands very clear-cut that the human reasoning is much more powerful than mechanical (TM) computation.

# 6. Baker-Gill-Solovay Theorem and the Proof

Verify that the proof does not use the diagonalization method (except in the justified special cases in Section 5) and it is based about the difference, on worst cases, between running times from a DTM and an NTM that recognize the $L_Z$-language ***L***, as demonstrated in Way 1.i of Section 5 compared to Section 4.

Moreover, notice that the addition into the proof methods of oracles to a PSPACE-Complete language *W* does not imply that false statement $P^W \neq NP^W$ (because the proof cannot be adapted to demonstrate that $P^W \neq NP^W$, since a DTM **Q** with an oracle to *W* could simulate any NTM with the same oracle using only a poly(n)-quantity of space, in an adapted Way 1.i, which would otherwise prove that $P^W = NP^W$).

These facts imply that the Baker-Gill-Solovay Theorem of inseparability of the classes P and NP by oracle-invariant methods (techniques that are conserved under the addition of oracles, like the pure diagonalization method without *algebraic oracle* [8]) does not refute this P ≠ NP proof. In other words, my proof technique does not *relativize* [4].

## 7. Razborov-Rudich Theorem and the Proof

SAT's weakness – The proof does not try to prove any lower bounds on the circuit complexity of a Boolean function, because it does not try to solve the still open question whether SAT is in P, since to prove P ≠ NP it was not necessary to solve the SAT question (for the proof, different from the wrong conclusion in [3, 6], it is irrelevant whether SAT is in P), whereas it was enough to prove that XG-SAT is in NP but not in P: Thus, the Razborov-Rudich Theorem of the Natural Proofs does not refute this proof. In other words, my proof technique does not *naturalize* [7].

## 8. Related Work, Aaronson-Wigderson Theorem and the Proof

There is no relevant related work on the goal to really solve the P versus NP question. From important papers upon the matter, there are only some "negative" results, like the ones referred to in Sections 6 and 7 and, more recently, as an extension of the *relativization* in Section 6, the proof that techniques that are conserved under the addition of an oracle and a low-degree extension of it over a finite field or ring cannot work on this question too, by the concept of *algebrization*, explained in [8].

Remember, however, that my proof does not use the pure diagonalization method (as referred to in Section 6), but it exploits properties of computation that are specific to real world computers, and then this new barrier is not valid to refute it, too.

## 9. Expert Advice & Academic Honesty

A reviewer, referring to the technical report in [15], has said "– It is disconcerting to see how the present author continues to ignore expert advice. His title borders on, and perhaps transgresses, academic honesty. Papers with such grandiose claims should only be considered after an endorsement by an expert."

The heart of my paper is just challenging some traditional definitions on TCS field, essentially the need of polynomial uniformity on the definitions of the complexity classes P and NP. However, that technical report says, for instance: "– As ... Definition 3.5 of his paper ... needs to before the universal quantification on *x* fix a polynomial bounding the length of the certificates, we from here on assume that his definition is viewed as being modified to do that ..."

So, as my proposed new definitions are so distorted in that expert advice, it has very low value in order to evaluate my proof, thus ignoring it is not really academic dishonesty at all, but only logical consequence of that challenge upon enhancing those definitions.

## 10. Freedom & Mathematics

"**– The essence of Mathematics is Freedom.**" (Georg Cantor) [22]

**André Luiz Barbosa – Goiânia - GO, Brazil – e-Mail: webmaster@andrebarbosa.eti.br – July 2009**

Site........ : www.andrebarbosa.eti.br
Blog........ : blog.andrebarbosa.eti.br

This Paper : www.andrebarbosa.eti.br/P_different_NP_Proof_Eng.htm
PDF....... : www.andrebarbosa.eti.br/P_different_NP_Proof_Eng.pdf
arXiv...... : http://arxiv.org/ftp/arxiv/papers/0907/0907.3965.pdf